\documentclass[twocolumn,aps,showpacs,superscriptaddress]{revtex4}
\usepackage{graphicx}
\usepackage{dcolumn}
\usepackage{bm}

\usepackage{color}

\begin{document}
\addtolength{\topmargin}{1.0in}

\def\NCA{Nuovo Cimento}
\def\NIM{Nucl. Instr. Meth.}
\def\NIMA{{Nucl. Instr. Meth.} A}
\def\NPB{{Nucl. Phys.} B}
\def\NPA{{Nucl. Phys.} A}
\def\PLB{{Phys. Lett.}  B}
\def\PRL{Phys. Rev. Lett.}
\def\PRC{{Phys. Rev.} C}
\def\PRD{{Phys. Rev.} D}
\def\ZPC{{Z. Phys.} C}
\def\JPG{{J. Phys.} G}
\def\CPC{Comput. Phys. Commun.}
\def\EPJ{{Eur. Phys. J.} C}

\preprint{}

\title{Partonic flow and $\phi$-meson production in Au+Au collisions at
    $\sqrt{s_{NN}}$ = 200 GeV}

\affiliation{Argonne National Laboratory, Argonne, Illinois 60439}
\affiliation{University of Birmingham, Birmingham, United Kingdom}
\affiliation{Brookhaven National Laboratory, Upton, New York 11973}
\affiliation{California Institute of Technology, Pasadena, California 91125}
\affiliation{University of California, Berkeley, California 94720}
\affiliation{University of California, Davis, California 95616}
\affiliation{University of California, Los Angeles, California 90095}
\affiliation{Carnegie Mellon University, Pittsburgh, Pennsylvania 15213}
\affiliation{University of Illinois at Chicago, Chicago, Illinois 60607}
\affiliation{Creighton University, Omaha, Nebraska 68178}
\affiliation{Nuclear Physics Institute AS CR, 250 68 \v{R}e\v{z}/Prague, Czech Republic}
\affiliation{Laboratory for High Energy (JINR), Dubna, Russia}
\affiliation{Particle Physics Laboratory (JINR), Dubna, Russia}
\affiliation{University of Frankfurt, Frankfurt, Germany}
\affiliation{Institute of Physics, Bhubaneswar 751005, India}
\affiliation{Indian Institute of Technology, Mumbai, India}
\affiliation{Indiana University, Bloomington, Indiana 47408}
\affiliation{Institut de Recherches Subatomiques, Strasbourg, France}
\affiliation{University of Jammu, Jammu 180001, India}
\affiliation{Kent State University, Kent, Ohio 44242}
\affiliation{Institute of Modern Physics, Lanzhou, China}
\affiliation{Lawrence Berkeley National Laboratory, Berkeley, California 94720}
\affiliation{Massachusetts Institute of Technology, Cambridge, MA 02139-4307}
\affiliation{Max-Planck-Institut f\"ur Physik, Munich, Germany}
\affiliation{Michigan State University, East Lansing, Michigan 48824}
\affiliation{Moscow Engineering Physics Institute, Moscow Russia}
\affiliation{City College of New York, New York City, New York 10031}
\affiliation{NIKHEF and Utrecht University, Amsterdam, The Netherlands}
\affiliation{Ohio State University, Columbus, Ohio 43210}
\affiliation{Panjab University, Chandigarh 160014, India}
\affiliation{Pennsylvania State University, University Park, Pennsylvania 16802}
\affiliation{Institute of High Energy Physics, Protvino, Russia}
\affiliation{Purdue University, West Lafayette, Indiana 47907}
\affiliation{Pusan National University, Pusan, Republic of Korea}
\affiliation{University of Rajasthan, Jaipur 302004, India}
\affiliation{Rice University, Houston, Texas 77251}
\affiliation{Universidade de Sao Paulo, Sao Paulo, Brazil}
\affiliation{University of Science \& Technology of China, Hefei 230026, China}
\affiliation{Shanghai Institute of Applied Physics, Shanghai 201800, China}
\affiliation{SUBATECH, Nantes, France}
\affiliation{Texas A\&M University, College Station, Texas 77843}
\affiliation{University of Texas, Austin, Texas 78712}
\affiliation{Tsinghua University, Beijing 100084, China}
\affiliation{Valparaiso University, Valparaiso, Indiana 46383}
\affiliation{Variable Energy Cyclotron Centre, Kolkata 700064, India}
\affiliation{Warsaw University of Technology, Warsaw, Poland}
\affiliation{University of Washington, Seattle, Washington 98195}
\affiliation{Wayne State University, Detroit, Michigan 48201}
\affiliation{Institute of Particle Physics, CCNU (HZNU), Wuhan 430079, China}
\affiliation{Yale University, New Haven, Connecticut 06520}
\affiliation{University of Zagreb, Zagreb, HR-10002, Croatia}

\author{B.I.~Abelev}\affiliation{University of Illinois at Chicago, Chicago, Illinois 60607}
\author{M.M.~Aggarwal}\affiliation{Panjab University, Chandigarh 160014, India}
\author{Z.~Ahammed}\affiliation{Variable Energy Cyclotron Centre, Kolkata 700064, India}
\author{B.D.~Anderson}\affiliation{Kent State University, Kent, Ohio 44242}
\author{D.~Arkhipkin}\affiliation{Particle Physics Laboratory (JINR), Dubna, Russia}
\author{G.S.~Averichev}\affiliation{Laboratory for High Energy (JINR), Dubna, Russia}
\author{Y.~Bai}\affiliation{NIKHEF and Utrecht University, Amsterdam, The Netherlands}
\author{J.~Balewski}\affiliation{Indiana University, Bloomington, Indiana 47408}
\author{O.~Barannikova}\affiliation{University of Illinois at Chicago, Chicago, Illinois 60607}
\author{L.S.~Barnby}\affiliation{University of Birmingham, Birmingham, United Kingdom}
% ...

\collaboration{STAR Collaboration}\noaffiliation

\date{\today}

\begin{abstract}

We present first measurements of the $\phi$-meson elliptic flow
($v_{2}(p_{T})$) and high statistics $p_{T}$ distributions for
different centralities from $\sqrt{s_{NN}}$ = 200 GeV Au+Au collisions
at RHIC. In minimum bias collisions the $v_{2}$ of the $\phi$ meson is
consistent with the trend observed for mesons. The ratio of the yields
of the $\Omega$ to those of the $\phi$ as a function of transverse
momentum is consistent with a model based on the recombination of
thermal $s$ quarks up to $p_{T}\sim 4$ GeV/$c$, but disagrees at
higher momenta. The nuclear modification factor ($R_{CP}$) of $\phi$
follows the trend observed in the $K^{0}_{S}$ mesons rather than in
$\Lambda$ baryons, supporting baryon-meson scaling. Since
$\phi$-mesons are made via coalescence of seemingly thermalized
$s$ quarks in central Au+Au collisions, the observations imply hot and dense matter with
partonic collectivity has been formed at RHIC.
\end{abstract}
\maketitle

The primary aim of ultra-relativistic heavy-ion collisions is to
produce and study a state of high-density nuclear matter called the
Quark-Gluon Plasma (QGP), the existence of which is supported by
lattice QCD calculations \cite{Karsch,Fodor2,QM04}. In the search for
this new form of matter, penetrating probes are essential in order to
gain information from the earliest stage of the
collisions. Phenomenological analysis~\cite{Shor} has suggested a
relatively small hadronic interaction cross section for $\phi$-mesons
although discussions about the $\phi-$proton interaction cross section
are yet to be conclusive~\cite{fp,fp2}. Therefore $\phi$-mesons
from high-energy nuclear collisions are expected to
provide information about the early partonic stages of the system's
evolution since they should remain mostly unaffected by
hadronic interactions. This is further supported by recent
measurements~\cite{STARphi200} which have ruled out the idea of
$\phi$-meson production by kaon coalescence.

Elliptic flow, $v_{2}$, is an observable which is thought to reflect
conditions from the early stage of the collision
\cite{v2Early1,v2Early2}. In non-central heavy-ion collisions, the
initial spatial anisotropy of the overlap region of the colliding
nuclei is transformed into an anisotropy in momentum space through
interactions between the particles. Systematic measurements of the
$v_{2}$ for the strange hadrons $K^{0}_{S}$, $\Lambda$, $\Xi$ and
$\Omega$ suggest that collectivity is developed at the partonic stage
at RHIC~\cite{KshortLamRCP,Omegav2}. The evidence for partonic
collectivity, one of the conditions for QGP formation, will be
further strengthened if it can be shown that $\phi$-mesons flow
like the other mesons.

A mass ordering predicted by hydrodynamics
\cite{Hydro,v2MultiS,Hirano} for $v_2$($p_T$) of identified particles
has been observed for $p_T \le 2$ GeV/c. At intermediate transverse
momentum, $2 \le p_T \le 5$ GeV/c, a separation of baryons and mesons
has been observed in measurements of both $v_2$ and the nuclear
modification factor,
$R_{CP}$~\cite{KshortLamRCP,PHENIXv2,KStarRCP}. These results are
consistent with calculations from quark recombination
models~\cite{VoloshinQM02,MolnarVoloshin,DukeRcp,Jinhui} implying the
deconfinement of the system prior to hadronization. The $\phi$ is a
vector meson, comparable in mass to the proton and $\Lambda$ baryons
with a relatively long lifetime. Its $v_2$ and $R_{CP}$ will provide a
critical test of the assumed underlying dynamics. In addition, as
argued in \cite{RudyHwa}, the ratio of the $\Omega$-baryon over
$\phi$-meson yields can be used to test the nature of light-quark
thermalization in the medium. The model predicts that the ratio of
the $\Omega$ to $\phi$ yields will rise monotonically.

The results presented in this paper were obtained with the STAR
detector \cite{STAR} at the Relativistic Heavy Ion Collider (RHIC) at
Brookhaven National Laboratory.  The detector components used in this
analysis were the Time Projection Chamber (TPC), and trigger
detectors, namely the zero degree calorimeters. 
Central collisions were selected using the Central Trigger Barrel. The collision
centrality was determined by the charged hadron multiplicity within pseudo-rapidity $|\eta|<0.5$.

High-statistics Au+Au data were taken in the 2004 run at
$\sqrt{s_{NN}}$ = 200 GeV. The minimum bias dataset used in this
analysis consisted of $\sim$13.5 million events and the
central-triggered dataset comprised about 10 million events.  Events
were required to have a primary vertex $z$ position (where $z$ is the
direction of the beam axis) within 30 cm of the center of the
TPC. Events from the minimum bias dataset were divided into 8
centrality bins: 0-10$\%$, 10-20$\%$, 20-30$\%$, 30-40$\%$, 40-50$\%$,
50-60$\%$, 60-70$\%$, and 70-80$\%$ of the measured cross-section. The
central-triggered dataset was used to extract the 0-5$\%$ and 0-12\%
data.

The $\phi$ yield in each $p_T$ bin was extracted from the invariant
mass ($m_{inv}$) distributions of $K^{+}+K^-$ candidates after
subtraction of combinatorial background estimated using event mixing
\cite{STARphi200}. The kaons were identified through their $dE/dx$ 
energy loss in the STAR TPC~\cite{STAR}.  Including the detector
resolution, the values of the reconstructed $\phi$ mass and width are
consistent with the PDG values~\cite{pdg}. The relative systematic
uncertainty due to the $dE/dx$ cut was estimated to be $\sim8\%$ by
using different cuts and comparing the yields after a particle
identification efficiency correction. 
Uncertainty in the residual background shape of the $m_{inv}$ distributions
resulted in a contribution of about 4.5\% to the errors on the final yields.

The $\phi$-meson $v_{2}$ results were obtained using the $v_{2}$ vs.
$m_{inv}$ method described in ref.~\cite{Borghini}. 
The method involves calculating the $v_{2}$ of the
same-event distribution as a function of $m_{inv}$ and then fitting
the resulting $v_{2}(m_{inv})$ distribution using:

\begin{equation}\label{eq:v2Tot}
v_{2}(m_{inv}) = v_{2S}\alpha(m_{inv}) + v_{2B}(m_{inv})[1-\alpha(m_{inv})]
\end{equation}

\noindent where $v_{2S}\equiv v_{2\phi}$ is the signal $v_{2}$ 
and $v_{2B}$ is the background $v_{2}$. $\alpha(m_{inv}) = S/(S+B)$ is
the ratio of the signal over the sum of the signal plus background of
the $m_{inv}$ distributions. It was extracted from fits (Breit-Wigner
plus a linear function) to the $\phi$ mass-peak for each $p_{T}$
bin. For each $p_{T}$ bin, the $v_{2}(m_{inv})$ was
fitted using Eq.~\ref{eq:v2Tot} in order to extract the fitting
parameter $v_{2S}$ and $v_{2B}$ was parameterized using a linear or
quadratic function in $m_{inv}$. These results are consistent with
results using an established method~\cite{v2Methods} where the
$\phi$-meson yield is plotted as a function of the difference between
its azimuthal angle and the estimated reaction plane angle,
$(\phi-\Psi)$. The values of $v_{2}$ are extracted from the fitting
to the function $dN/d\phi = P_{0}(1 + 2v_{2}\cos(2(\phi-\Psi)))$.

\begin{figure}[!hbt]
  \center
  \includegraphics[width=0.4\textwidth,height=0.45\textheight]{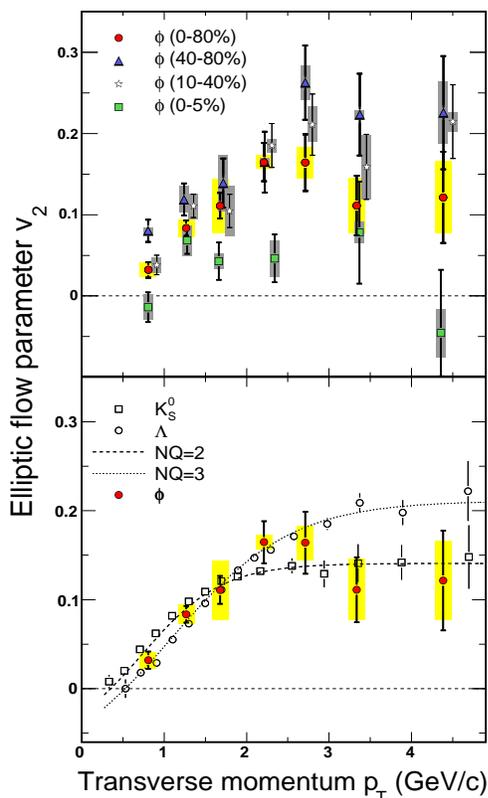}
  \vspace{-0.45cm}  
  \caption{\label{fig:V2}(color online) \textbf{Top panel:} The elliptic flow,
    $v_{2}(p_{T})$, for the $\phi$-meson as a function of centrality. 
    The vertical
    error bars represent the statistical errors while the shaded bands
    represent the systematic uncertainties. For clarity, data points are shifted slightly.
    \textbf{Bottom panel:} Minimum bias $v_{2}(p_{T})$ for the
    $\phi$-meson compared to results for $\Lambda$ and $K^{0}_{S}$~\cite{KshortLamRCP}.
    The dashed and dotted lines represent parameterizations
    inspired by number-of-quark scaling ideas from ref.~\cite{NCQ} for NQ=2 and NQ=3
    respectively.} 
\end{figure}

In the top panel of Fig.~\ref{fig:V2} we present the first measurement
of the differential elliptic flow, $v_{2}(p_{T})$, of the $\phi$-meson
from Au+Au collisions for four centrality bins.  In this and the
following figures, the vertical error bars on the $\phi$ data points
indicate the statistical errors while the shaded bands indicate the
extent of the systematic uncertainties.  The systematic errors vary from point to 
point including uncertainties in extracting the signal for 
obtaining $\alpha(m_{inv})$ and differences in the reaction plane resolution determination.
For minimum bias collisions, an additional contribution to account for the different methods of 
extracting the $v_2$($p_T$) values is also included in the systematic error.
Non-flow effects~\cite{starnf,FlowPRC} are not included in the
systematic error. As expected, $v_{2}(p_{T})$ increases with
increasing eccentricity (decreasing centrality) of the initial overlap region. This
trend is also illustrated in Table~\ref{table1} which presents the
$p_T$-integrated values of $\phi$-meson elliptic flow, $\langle
v_{2}\rangle$, calculated by convoluting the $v_{2}(p_{T})$ with the
respective $p_{T}$ spectrum for three centrality bins. It should be
noted that the centrality dependence of the $\langle v_{2}\rangle$ of
$\phi$-mesons is consistent with that of charged
hadrons~\cite{FlowPRC}.

\begin{table}[h!]
\centering \caption{ Integrated elliptic flow, $\langle v_{2}\rangle$,
  for the $\phi$-meson for three centrality bins.} \label{table1}
\begin{ruledtabular}
\begin{tabular}{cc} Centrality (\%) & $\langle v_{2}\rangle$ (\%)  \\ \hline
40 -- 80  &  8.5 $\pm ^{1.1 (stat)}_{0.2 (sys)}$  \\ 
10 -- 40  &  6.6 $\pm ^{0.8 (stat)}_{0.2 (sys)}$  \\
0 -- 5    &  2.1 $\pm ^{1.2 (stat)}_{0.5 (sys)}$  \\
\end{tabular}
\end{ruledtabular}
\end{table}

The lower panel of Fig.~\ref{fig:V2} shows the minimum bias (0-80\%)
result compared to parameterizations for number-of-quark scaling for
mesons (NQ=2) and baryons (NQ=3) whose free parameters have been fixed
by fitting to the $\Lambda$ and $K^{0}_{S}$ results
simultaneously~\cite{NCQ}. In this case, for $p_{T} <$ 2 GeV/$c$, the
$\phi$ $v_{2}$ follows a mass-ordered hierarchy where the values of
$v_{2}$, within errors, fall between those of the heavier $\Lambda$
(open circles) and lighter $K^{0}_{S}$ (open-squares). However, at
intermediate $p_{T}$, between 2-5 GeV/$c$, the $\phi$ $v_{2}$ appears
to follow the same trend as $K_S^0$. When we fit the $v_{2}$($p_T$) of
$\phi$-mesons with the quark number scaling ansatz~\cite{NCQ}, the
resulting fit parameter NQ $=2.3\pm0.4$. The fact that the $\phi$
$v_{2}(p_{T})$ is the same as that of other mesons indicates that the
heavier $s$ quarks flow as strongly as the lighter $u$ and $d$
quarks. As previously mentioned, $\phi$-mesons are not formed through
kaon coalescence~\cite{STARphi200} and do not participate strongly in
hadronic interactions. Therefore the results demonstrate partonic
collectivity.

\begin{figure}[!hbt]
\vspace{-0.45cm}
  \center
  \includegraphics[scale=0.35]{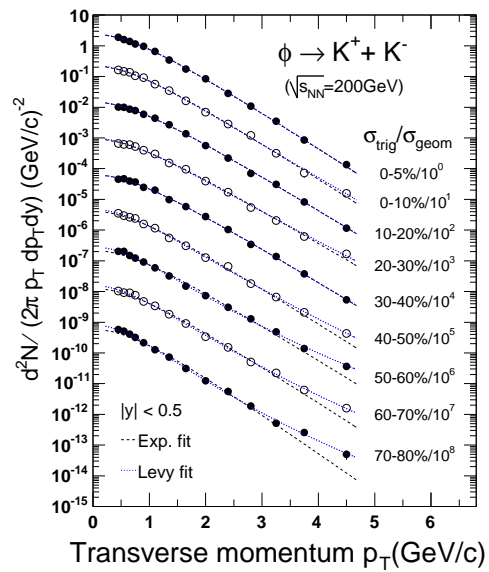}
\vspace{-0.55cm}
  \caption{\label{fig:Spectra}(color online) Transverse momentum distributions of $\phi$-mesons
    from Au+Au collisions at $\sqrt{s_{NN}}$ = 200 GeV. For clarity, distributions for
    different centralities are scaled by factors of ten. Dashed lines represent
    the exponential fits to the distributions and the dotted lines are
    Levy function fits. Error bars represent statistical
    errors only.}
\end{figure}

Figure~\ref{fig:Spectra} shows the $p_{T}$ distributions of
$\phi$-mesons as a function of centrality. The central-triggered
dataset was used to obtain the most central spectrum while the other
distributions were obtained using the minimum bias dataset. The error
bars shown in Fig.~\ref{fig:Spectra} are statistical only. In the figure, 
the errors are smaller than the size of the data points.

Each $p_{T}$ spectrum in Fig.~\ref{fig:Spectra} has been fitted using
both an exponential function (dashed lines) in $m_T$ and a Levy function
(dotted lines) which has an exponential-like shape at low $p_{T}$ and
is power-law-like at higher $p_{T}$. While the central data are fitted
equally well by both functions the more peripheral spectra are better
fitted by the Levy function indicating less thermal contributions in
peripheral collisions.

\begin{figure}[!hbt]
  \center
  \includegraphics[scale=0.35]{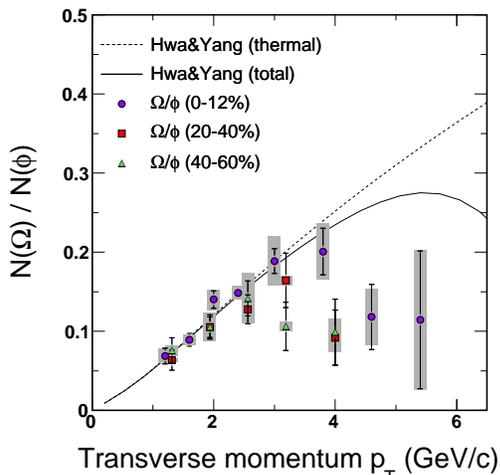}
\vspace{-0.5cm}
  \caption{\label{fig:OmegaPhiRatio}(color online) The $N(\Omega)/N(\phi)$ ratio
  vs.\ $p_{T}$ for three centrality bins in $\sqrt{s_{NN}}$ = 200 GeV
  Au+Au collisions. The solid and dashed lines represent
  recombination model predictions for central collisions~\cite{RudyHwa} for total and thermal
  contributions, respectively.}
\end{figure}  

In Fig.~\ref{fig:OmegaPhiRatio}, the ratios of $N(\Omega)/N(\phi)$
vs.\ $p_{T}$ are presented as a function of centrality. The $\Omega$
datapoints are from ref.~\cite{STARWhiteP} (for 0-10\%) and
ref.~\cite{ScalingHyperon} for the other centralities. The errors of
the ratios are dominated by the $\Omega$ datapoints. Also shown in the
figure are recombination model expectations for central
collisions~\cite{RudyHwa} based on $\phi$ and $\Omega$ production from
coalescence of thermal $s$ quarks in the medium. The model describes
the trend of the data up to $p_{T}\sim4$ GeV/c but fails at higher
$p_{T}$. Other models based on dynamical recombination of
quarks~\cite{DukeRcp, TexasPhi} were also compared to the
data. However, ref.~\cite{DukeRcp} overpredicts the ratio while
ref.~\cite{TexasPhi} gives the wrong shape. With decreasing centrality, the
observed $N(\Omega)/N(\phi)$ ratios seem to turn over at successively
lower values of $p_{T}$ indicating a smaller contribution from thermal
quark coalescence in more peripheral collisions. This is also
reflected in the smooth evolution of the spectra shapes from the
thermal-like exponential to power-law shapes shown in
Fig.~\ref{fig:Spectra}.

\begin{figure}[!hbt]
  \center
  \includegraphics[scale=0.38]{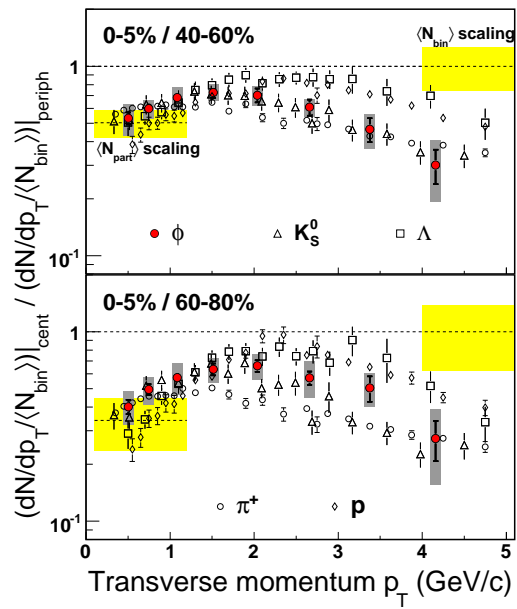}
  \vspace{-0.45cm}
  \caption{\label{fig:Rcp}(color online) The $R_{CP}$ of mid-rapidity
   $\phi$-mesons produced in $\sqrt{s_{NN}}$ = 200 GeV Au+Au collisions: 
   (top) 0-5$\%$ vs.\ 40-60$\%$ and (bottom) 0-5$\%$ vs.\ 60-80$\%$
  The shaded bands
  represent the uncertainties in the Glauber model calculations for
  $\langle N_{bin}\rangle$ and $\langle
  N_{part}\rangle$~\cite{Glauber}. Also shown are results for
  $\Lambda$ and $K^{0}_{S}$~\cite{KshortLamRCP} and protons and
  $\pi^{+}$~\cite{PPi}.}
\end{figure}

The nuclear modification factor $R_{CP}(p_{T})$ measures the change of
$p_{T}$ distributions from peripheral to central collisions and has
been measured for most of the identified hadrons. In
Fig.~\ref{fig:Rcp}, the high statistics $\phi$-meson $R_{CP}$ (solid
circles) is compared to $K^{0}_{S}$ (open triangles) and $\Lambda$ (open
squares) from ref.~\cite{KshortLamRCP} for two different centrality
combinations (upper and lower panels). In both panels the binary-scaled yield of
$\phi$-mesons is suppressed ($R_{CP}$ below unity) in central compared
to peripheral collisions. It has been shown that $u$, $d$ and
$s$ quarks are approaching equilibration at hadronization~\cite{STARWhiteP}. 
The  $\phi$-meson $R_{CP}$ is more consistent with that of $K^{0}_{S}$ (meson) than of $\Lambda$
(baryon) for the 0-5\%/40-60\% case (upper panel).
This is as predicted by particle production models based on
recombination of thermal quarks~\cite{DukeRcp}. For the more peripheral bin
(lower panel) the $\phi$ $R_{CP}$ falls between that of the
$\Lambda$ and $K^{0}_{S}$. In the 60-80\% centrality bin (see lower, binary 
collision-scaled $\phi$ production is very similar to that in p+p and d+Au
collisions where strangeness production is canonically
suppressed~\cite{cs}. Therefore a baryon-meson scaling behaviour of $R_{CP}$ is
not expected in the lower panel of Fig.~\ref{fig:Rcp}. In addition,
for baryons and mesons respectively, there seems to be an ordering in terms of strangeness content. 
This has also been observed in $R_{AA}$ for strange particles~\cite{Sevil}.

In summary, we have presented first measurements of the elliptic flow
of $\phi$-mesons as a function of collision centrality in Au+Au
collisions at $\sqrt{s_{NN}}$ = 200 GeV.  At low $p_{T}$ ($<$ 2
GeV/$c$), $v_{2}$ is consistent with hydrodynamical expectations. At
intermediate $p_{T}$ (2 $<p_{T}<$ 5 GeV/$c$), $v_{2}$ of $\phi$-mesons
is consistent with number-of-quark scaling for mesons.  These
observations indicate the development of partonic collectivity in the
medium.  Measurements of the $\phi$ $p_{T}$ spectra as a function of
centrality show an evolution of the spectral shape from exponential to
power-law-like with decreasing centrality, reflecting the increasing
contributions from hard and possibly other non-equilibrium processes
in more peripheral collisions. The result of a recombination
model~\cite{RudyHwa} is consistent with the trend of the central
$N(\Omega)/N(\phi)$ ratio up to $p_{T}\sim4$ GeV/$c$ which covers more
than 95\% of the hadron yields. At higher $p_{T}$, the model
fails. The $\phi$-meson $R_{CP}$ resembles the $K^{0}_{S}$ for the
0-5\%/40-60\% case which is consistent with meson scaling.
Since $\phi$-mesons are made via coalescence of seemingly thermalized
$s$ quarks in central Au+Au collisions, the observations imply hot and dense matter with
partonic collectivity has been formed at RHIC. 

We thank the RHIC Operations Group and RCF at BNL, and the
NERSC Center at LBNL for their support. This work was supported
in part by the Offices of NP and HEP within the U.S. DOE Office 
of Science; the U.S. NSF; the BMBF of Germany; CNRS/IN2P3, RA, RPL, and
EMN of France; EPSRC of the United Kingdom; FAPESP of Brazil;
the Russian Ministry of Science and Technology; the Ministry of
Education and the NNSFC of China; IRP and GA of the Czech Republic,
FOM of the Netherlands, DAE, DST, and CSIR of the Government
of India; Swiss NSF; the Polish State Committee for Scientific 
Research; SRDA of Slovakia, and the Korea Sci. $\&$ Eng. Foundation. 
HGR thanks the Alexander von Humboldt Foundation for generous support.

\end{document}